\documentclass[cameraready]{Interspeech}

\title{Decoding Speech Envelopes from Electroencephalogram with a Contrastive Pearson Correlation Coefficient Loss}

\author[affiliation={1}, equalcontribution]{Yayun}{Liang}
\author[affiliation={1}, equalcontribution]{Yuanming}{Zhang}
\author[affiliation={2}]{Fei}{Chen}
\author[affiliation={1,3}]{Jing}{Lu}
\author[affiliation={1}]{Zhibin}{Lin}

\address{
$^1$ Key Laboratory of Modern Acoustics, Nanjing University, Nanjing 210093, China\\
$^2$ Department of Electrical and Electronic Engineering, Southern University of Science and Technology, Shenzhen 518055, China\\
$^3$ Nanjing University--Horizon Intelligent Audio Laboratory, Nanjing Institute of Advanced Artificial Intelligence, Nanjing, China
}

\email{yayunliang@smail.nju.edu.cn, yuanming.zhang@smail.nju.edu.cn, fchen@sustech.edu.cn,  lujing@nju.edu.cn, zblin@nju.edu.cn}


\usepackage{comment}
\usepackage{booktabs}   
\usepackage{multirow}   
\usepackage{makecell}   
\usepackage{graphicx}   
\usepackage{subcaption}
\usepackage{tikz}
\usepackage{makecell}
\begin{document}

\maketitle

\begin{abstract}
    
Recent advances in reconstructing speech envelopes from Electroencephalogram (EEG) signals have enabled continuous auditory attention decoding (AAD) in multi-speaker environments. Most Deep Neural Network (DNN)-based envelope reconstruction models are trained to maximize the Pearson correlation coefficients (PCC) between the attended envelope and the reconstructed envelope (attended PCC). While the difference between the attended PCC and the unattended PCC plays an essential role in auditory attention decoding, existing methods often focus on maximizing the attended PCC. We therefore propose a contrastive PCC loss which represents the difference between the attended PCC and the unattended PCC. The proposed approach is evaluated on three public EEG AAD datasets using four DNN architectures. Across many settings, the proposed objective improves envelope separability and AAD accuracy, while also revealing dataset- and architecture-dependent failure cases.

\end{abstract}

\section{Introduction}

Humans can selectively attend to a certain speaker (the target speaker) while suppressing competing speech. Recent studies in neuroscience have demonstrated that auditory selective attention enhances cortical tracking of the attended speech envelope \cite{ding2012emergence,mesgarani2012selective,Fan2025ListenNet,Zhang2025TimeFrequencyAAD}. Electroencephalogram (EEG)-based decoding has gained traction for AAD and brain-assisted target speaker extraction due to its noninvasiveness and practical setup requirements\cite{Fan2025M3ANet,Si2025TFGANet,Pan2024NeuroHeed, Hjortkjaer2025Realtime}.

Recent research has increasingly shifted toward Deep Neural Networks (DNNs), applied either to continuous speech-feature reconstruction \cite{accou2023decoding,fan2025ssm2mel,borges2025speech} or to end-to-end selective auditory attention decoding \cite{zhang2023learnable,xu2024densenet,ding2024eeg}. Most DNN-based envelope reconstruction models aim to maximize the Pearson correlation coefficient (PCC) between the attended envelope and the reconstructed envelope (attended PCC) \cite{accou2023decoding,fan2025ssm2mel}. However, maximizing the PCC with the attended speech does not explicitly prevent the model from remembering speech features common to both attended and unattended speakers, resulting in potentially decreased discrepancy between the attended and unattended PCC. 

To address this issue, we propose a contrastive PCC loss which represents the difference between the attended PCC and the unattended PCC, designed to explicitly enhance the separation between competing speech streams. We compare four state-of-the-art (SOTA) DNN models across three benchmark datasets between the commonly used PCC loss and the proposed contrastive PCC loss. 

Unlike prior work that emphasizes architectural innovations, we focuses on the design of regression objectives and their relationship to AAD decision metrics. The main contributions of this paper are summarized as follows:
\begin{itemize}
    \item We analyze the correlation between decoding accuracy and the difference between the attended PCC and the unattended PCC.
    \item We propose a contrastive Pearson correlation coefficient loss to address the existing issue that maximizing the PCC with the attended speech may also increase the PCC with the unattended speech.
    \item We validated the approach across three public datasets and four DNN architectures.
\end{itemize}

The remainder of this paper is organized as follows: 
Section~II describes the methods adopted in this work.
Section~III presents the experimental setup and processing pipeline.
Section~IV provides the results and discussion.
Finally, Section~V concludes the paper.

\section{Methods}

\subsection{Task description}

In this work, we formulate EEG-based speech envelope reconstruction -- a common AAD pipeline component --as a regression problem. Let $\mathbf{X} \in \mathbb{R}^{T \times C}$ denote the multichannel EEG input, where $T$ is the number of temporal samples and $C$ is the number of electrodes. We abstract the decoding model as a nonlinear mapping $f: \mathbb{R}^{T \times C} \rightarrow \mathbb{R}^{T\times1}$, which transforms input EEG segment $\mathbf{X}$ into the predicted speech envelope $\hat{\mathbf{y}} \in \mathbb{R}^{T\times1}$.

The acoustic reference is represented as $\mathbf{S} \in \mathbb{R}^{T \times N}$, where $N$ denotes the number of concurrent speakers in the auditory scene. The PCC between two signals $\mathbf{x}$ and $\mathbf{z}$ is defined as
\begin{equation}
\mathrm{corr}(\mathbf{x}, \mathbf{z})
=
\frac{
\sum_{t} (x_t - \bar{x})(z_t - \bar{z})
}{
\sqrt{\sum_{t} (x_t - \bar{x})^2}
\sqrt{\sum_{t} (z_t - \bar{z})^2}
},
\end{equation}
where $\bar{x}$ and $\bar{z}$ are the sample means of $\mathbf{x}$ and $\mathbf{z}$, respectively. Let the attended PCC $\rho_a$ be the Pearson correlation coefficient (PCC) between the attended envelope and the reconstructed envelope:
\[
\rho_a = \mathrm{corr}(\hat{\mathbf{y}}, \mathbf{S}_a),
\]
where the subscript \(a\) denotes the attended speech envelope.
Similarly, we define the correlation between the prediction and all unattended envelopes as
\[
\rho_{u,j} = \mathrm{corr}(\hat{\mathbf{y}}, \mathbf{S}_{u,j}), 
\qquad j = 1,\ldots, N-1,
\]
where \(u\) represents the unattended speech, with \(j\) indicating the unattended envelope index. The decoding accuracy is defined as
\begin{equation}
    Acc = \frac{1}{N_{sample}}
    \sum_{i=1}^{N_{sample}}
    \left[\prod_{j=1}^{N_{speaker}-1}(\rho_{a}(i)>\rho_{u,j}(i))\right],
\end{equation}
where $i$ is the sample index.

\subsection{Model Architectures}
\subsubsection{CNN-based Models}

VLAAI \cite{accou2023decoding} predicts continuous attended speech envelopes using stacked 1-D convolutions with skip connections. LSM-CNN \cite{zhang2023learnable} introduces a learnable spatial-mapping layer that converts multichannel EEG signal from 2-D to 3-D layout (two channel dimensions), enabling the representation of channel spatial locations and adjacency.

\subsubsection{Transformer-like models}

SSM2Mel \cite{fan2025ssm2mel} combines selective state-space Mamba blocks with multi-head self-attention, enabling efficient long-range EEG modeling for mel-spectrogram reconstruction. EEG-Deformer \cite{ding2024eeg} uses a hybrid CNN–Transformer architecture in which shallow and deep representations are produced with fine- and coarse-grain feature extractors. A sequence-to-sequence EEG-Deformer was adopted based on the original EEG-Deformer classification model. 

\subsection{Loss Function}

A common objective in speech-envelope reconstruction is to maximize the Pearson correlation coefficient between the predicted envelope $\hat{\mathbf{y}}$ and the attended envelope $S_a$.
The PCC loss is defined as
\begin{equation}
\mathcal{L}_{\mathrm{PCC}}
= -\rho_{a} = - \mathrm{corr}\left(\hat{\mathbf{y}}, \mathbf{S}_a\right),
\end{equation}
which maximizes the correlation between the reconstructed and attended envelopes. However, the difference between the attended PCC and the unattended PCC plays an essential role in auditory attention decoding, as identification of the attended envelope depends on high contrast between attended and unattended PCC. 

The proposed contrastive correlation objective explicitly separates attended and unattended envelopes. In practice, we found that naively maximizing 
$-\rho_{a} + \sum(\rho_{u,j})$ is fundamentally ill-posed: the model can increase the objective by driving all correlations toward negative values, yielding a numerically better loss without learning any meaningful auditory representation. To prevent the model from exploiting this shortcut, the mean of unattended PCCs is used in the proposed objective. The proposed $\mathcal{L}_{\Delta\mathrm{PCC}}$ maximizes the attended correlation while minimizing the mean correlation with unattended streams:
\begin{equation}
    \mathcal{L}_{\Delta\mathrm{PCC}}
    = -\rho_a 
      + \frac{1}{N_{speaker}-1}\sum_{j=1}^{N_{speaker}-1} \rho_{u,j} .
\end{equation}
By minimizing $\mathcal{L}_{\Delta\mathrm{PCC}}$, DNN models are enabled with enhanced attention discriminability and mitigated ambiguity inherent to the commonly-used PCC objective. 

\begin{figure*}[t]
    \centering

    \begin{tikzpicture}
        \node[inner sep=0] (imgA) {
            \includegraphics[width=0.45\textwidth]{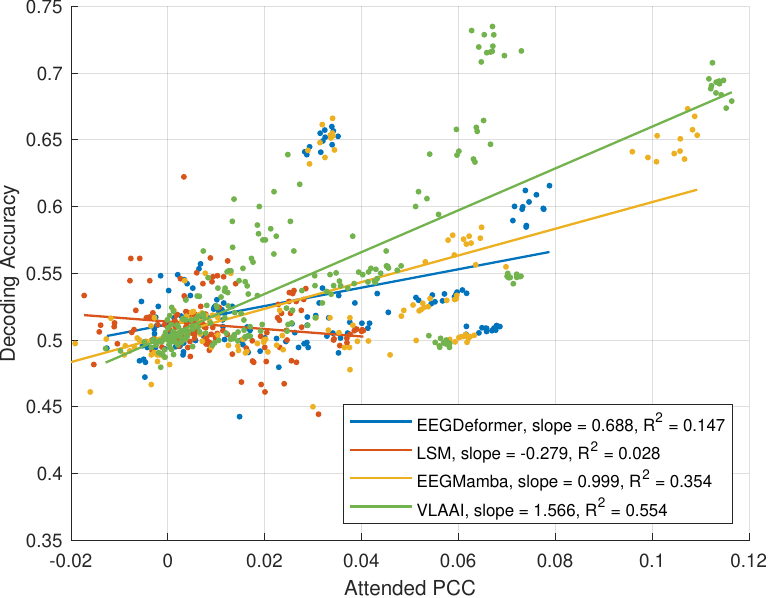}
        };
        \node[anchor=north west, font=\bfseries, xshift=-7pt, yshift=4pt] 
            at (imgA.north west) {(a)};
    \end{tikzpicture}
    \hfill
    \begin{tikzpicture}
        \node[inner sep=0] (imgB) {
            \includegraphics[width=0.45\textwidth]{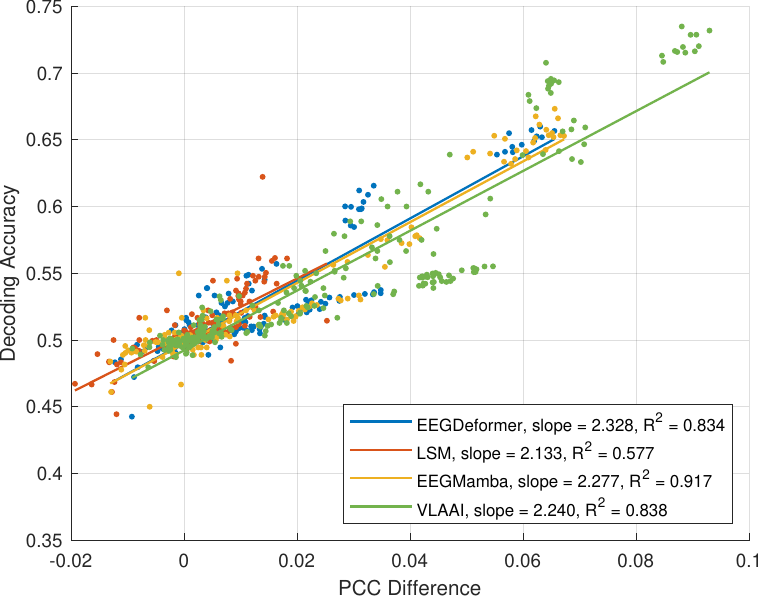}
        };
        \node[anchor=north west, font=\bfseries, xshift=-7pt, yshift=4pt] 
            at (imgB.north west) {(b)};
    \end{tikzpicture}

    \caption{The relationships between decoding accuracy (ACC) and (a) attended PCC,  (b) PCC difference between the attended PCC and the unattended PCC ($\Delta\mathrm{PCC}$). Each point represents the decoding accuracy of a DNN decoder trained on a specific dataset using either $\mathcal{L}_{\mathrm{PCC}}$ or $\mathcal{L}_{\Delta\mathrm{PCC}}$. Each line represents a linear trend fitted on results of a DNN decoder.}
    \label{fig:acc_relations}
\end{figure*}

\section{Experiment}

\subsection{Dataset}
We evaluate our methods on three publicly available EEG-based auditory attention datasets, covering two-talker paradigms: KUL \cite{biesmans2016auditory}, DTU \cite{fuglsang2017noise} and KUL-AV-GC \cite{rotaru2024we}.

The KUL dataset follows the auditory-attention paradigm described in \cite{biesmans2016auditory}. Sixteen normal-hearing subjects listened to two stories presented at $\pm90^\circ$, each story consisting of an attended and an unattended track with four parts of approximately six minutes; in this work, we use the whole dataset.

The DTU dataset is based on \cite{fuglsang2017noise}. Eighteen normal-hearing subjects listened to two narrated stories segmented into 50-second excerpts, presented simultaneously from $\pm60^\circ$ azimuth.

The KUL-AV-GC dataset corresponds to the audiovisual spatial-attention paradigm described in \cite{rotaru2024we}. Sixteen normal-hearing participants completed eight 10-minute trials, with two spatially separated speech streams rendered at $\pm90^\circ$ using individualized head-related impulse responses.

For these datasets, EEG was recorded using a 64-channel BioSemi ActiveTwo system.

\subsection{Data Preprocessing}

All datasets were preprocessed with the same pipeline to ensure a unified format for model training. EEG signals were first band-pass filtered between 1 and 32~Hz to retain low-frequency components associated with cortical speech tracking. After filtering and referencing, all EEG recordings were resampled to a common sampling rate of 128~Hz. Ocular and muscular artifacts were mitigated using Independent Component Analysis (ICA) and electrooculogram (EOG) regression when available, and bad channels were removed and interpolated. 

Speech envelopes for all talkers were extracted from the audio waveform using an ERB gammatone filterbank with 17 subbands spanning the frequency range from 50 to 5000~Hz.
The subband envelopes were obtained by magnitude extraction followed by a power-law nonlinearity with an exponent of 0.6, and were linearly summed to form a single broadband speech envelope.
The resulting speech envelopes were resampled to 128~Hz to match the EEG sampling rate.
EEG signals and speech envelopes were temporally aligned and segmented into matrices of size $T \times C$ and $T \times N$, respectively.

\subsection{Training Procedure}
All models were trained using a four fold leave-one-trial-out cross-validation scheme. This strategy ensured that no training, validation, or test segments originated from the same trial and prevented the model from utilizing trial fingerprints to obtain overestimated decoding results.

Each of the four models—VLAAI, LSM, EEGMamba, and EEGDeformer—was trained independently using both loss functions (the $\mathcal{L}_{\mathrm{PCC}}$ and the proposed $\mathcal{L}_{\Delta\mathrm{PCC}}$) in three datasets.Model parameters were optimized using the AdamW optimizer with an initial learning rate of $5\times10^{-4}$ and a weight decay of $5\times10^{-4}$.
Training was performed with a batch size of 64 for up to 100 epochs, and early stopping was applied when the validation loss did not improve for 10 consecutive epochs.
All experiments were conducted on a single NVIDIA RTX 5070~Ti GPU.
Code and trained models will be made publicly available upon acceptance.

\section{Results and Discussion}

\begin{table*}[t]
\centering
\scriptsize
\caption{
\textbf{Decoding accuracy, $\Delta$PCC, attended PCC and unattended PCC comparison across models, loss functions, datasets, and window lengths.}
        Decoding accuracy is defined as the proportion in which the attended PCC exceeded the unattended PCC.
        Attended PCC and unattended PCC denote the Pearson correlation coefficients between the reconstructed speech envelope and the attended and unattended speech envelopes, respectively. 
        $\Delta$PCC corresponds to the difference between attended and unattended PCC.
        Asterisks indicate statistical significance based on paired hypothesis testing
        ($^* p < 0.05$, $^{**} p < 0.01$, $^{***} p < 0.001$).
}
\label{tab:acc_pcc_merge}

\setlength{\tabcolsep}{4pt}
\renewcommand{\arraystretch}{1.22}

\begin{tabular*}{\textwidth}{@{\extracolsep{\fill}} c c c ccc ccc}
\toprule
\multirow{2}{*}{\textbf{Metric}} &
\multirow{2}{*}{\textbf{Model}} &
\multirow{2}{*}{\textbf{Loss}} &
\multicolumn{3}{c}{\textbf{1 s}} &
\multicolumn{3}{c}{\textbf{10 s}} \\
\cmidrule(lr){4-6} \cmidrule(lr){7-9}
& & &
KUL & DTU & KUL-AV-GC &
KUL & DTU & KUL-AV-GC \\
\midrule

\multirow{8}{*}{Accuracy}

& \multirow{2}{*}{VLAAI}
& $\mathcal{L}_{\mathrm{PCC}}$
& 0.5474 & 0.5284 & 0.5106
& 0.6899 & 0.6445 & 0.5519 \\
& 
& $\mathcal{L}_{\Delta\mathrm{PCC}}$
& 0.5483 & 0.5428$^{***}$ & 0.5147
& 0.7209$^{***}$ & 0.6394 & 0.583$^{*}$ \\

& \multirow{2}{*}{LSM}
& $\mathcal{L}_{\mathrm{PCC}}$
& 0.5013 & 0.5037 & 0.4901
& 0.5043 & 0.5271 & 0.4812 \\
& 
& $\mathcal{L}_{\Delta\mathrm{PCC}}$
& 0.5039 & 0.5086$^{**}$ & 0.5090$^{***}$
& 0.5136 & 0.5448$^{**}$ & 0.5296$^{***}$ \\

& \multirow{2}{*}{EEGMamba}
& $\mathcal{L}_{\mathrm{PCC}}$
& 0.5268 & 0.5013 & 0.4948
& 0.6486 & 0.5704 & 0.5040 \\
&
& $\mathcal{L}_{\Delta\mathrm{PCC}}$
& 0.5181 & 0.4994 & 0.4994
& 0.6494$^{***}$ & 0.5012 & 0.5179 \\

& \multirow{2}{*}{EEGDeformer}
& $\mathcal{L}_{\mathrm{PCC}}$
& 0.5325 & 0.5081 & 0.5006
& 0.5995 & 0.5097 & 0.4951 \\
&
& $\mathcal{L}_{\Delta\mathrm{PCC}}$
& 0.5225 & 0.4999 & 0.5033
& 0.6494$^{***}$ & 0.5341$^{**}$ & 0.5099 \\

\midrule

\multirow{8}{*}{$\Delta$PCC}

& \multirow{2}{*}{VLAAI}
& $\mathcal{L}_{\mathrm{PCC}}$
& 0.0436 & 0.0323 & 0.0130
& 0.0641 & 0.0678 & 0.0226 \\
&
& $\mathcal{L}_{\Delta\mathrm{PCC}}$
& 0.0479$^{***}$ & 0.0428$^{***}$ & 0.0168
& 0.0887$^{***}$ & 0.0652 & 0.0302$^{*}$ \\

& \multirow{2}{*}{LSM}
& $\mathcal{L}_{\mathrm{PCC}}$
& 0.0013 & 0.0029 & -0.0071
& 0.0016 & 0.0095 & -0.0109 \\
&
& $\mathcal{L}_{\Delta\mathrm{PCC}}$
& 0.0033 & 0.0071$^{**}$ & 0.0075$^{***}$
& 0.0051 & 0.0135$^{**}$ & 0.0070$^{***}$ \\

& \multirow{2}{*}{EEGMamba}
& $\mathcal{L}_{\mathrm{PCC}}$
& 0.0247 & 0.0026 & -0.0048
& 0.0583 & 0.0364 & 0.0019 \\
&
& $\mathcal{L}_{\Delta\mathrm{PCC}}$
& 0.0176 & 0.0002 & -0.0017
& 0.0633$^{***}$ & 0.0001 & 0.0046 \\

& \multirow{2}{*}{EEGDeformer}
& $\mathcal{L}_{\mathrm{PCC}}$
& 0.0295 & 0.0103 & 0.0002
& 0.0307 & 0.0027 & -0.0004 \\
&
& $\mathcal{L}_{\Delta\mathrm{PCC}}$
& 0.0205 & 0.0000 & 0.0040
& 0.0603$^{***}$ & 0.0102$^{**}$ & 0.0012 \\
\midrule
\multirow{8}{*}{\makecell{attended\\PCC}}

& \multirow{2}{*}{VLAAI}
& $\mathcal{L}_{\mathrm{PCC}}$
& 0.0712 & 0.0742 & 0.0202
& 0.1136 & 0.0805 & 0.0361 \\
&
& $\mathcal{L}_{\Delta\mathrm{PCC}}$
& 0.0405 & 0.0366 & 0.0088
& 0.0666 & 0.0607 & 0.0170 \\

& \multirow{2}{*}{LSM}
& $\mathcal{L}_{\mathrm{PCC}}$
& 0.0095 & 0.0379 & 0.0189
& 0.0061 & 0.0256 & 0.0215 \\
&
& $\mathcal{L}_{\Delta\mathrm{PCC}}$
& 0.0024 & 0.0038 & -0.0055
& 0.0012 & 0.0090 & -0.0030 \\

& \multirow{2}{*}{EEGMamba}
& $\mathcal{L}_{\mathrm{PCC}}$
& 0.0544 & 0.0607 & 0.0182
& 0.1044 & 0.0616 & 0.0360 \\
&
& $\mathcal{L}_{\Delta\mathrm{PCC}}$
& 0.0061 & 0.0033 & -0.0013
& 0.0326 & -0.0017 & 0.0076 \\

& \multirow{2}{*}{EEGDeformer}
& $\mathcal{L}_{\mathrm{PCC}}$
& 0.0568 & 0.0656 & 0.0220
& 0.0746 & 0.0386 & 0.0290 \\
&
& $\mathcal{L}_{\Delta\mathrm{PCC}}$
& 0.0108 & 0.0045 & -0.0002
& 0.0319 & 0.0010 & 0.0008 \\
\midrule

\multirow{8}{*}{\makecell{unattended\\PCC}}

& \multirow{2}{*}{VLAAI}
& $\mathcal{L}_{\mathrm{PCC}}$
& 0.0277 & 0.0420 & 0.0072
& 0.0495 & 0.0127 & 0.0134 \\
&
& $\mathcal{L}_{\Delta\mathrm{PCC}}$
& -0.0073 & -0.0061 & -0.0080
& -0.0221 & -0.0045 & -0.0132 \\

& \multirow{2}{*}{LSM}
& $\mathcal{L}_{\mathrm{PCC}}$
& 0.0082 & 0.0350 & 0.0260
& 0.0045 & 0.0161 & 0.0325 \\
&
& $\mathcal{L}_{\Delta\mathrm{PCC}}$
& -0.0009 & -0.0033 & -0.0129
& -0.0039 & -0.0045 & -0.0099 \\

& \multirow{2}{*}{EEGMamba}
& $\mathcal{L}_{\mathrm{PCC}}$
& 0.0296 & 0.0581 & 0.0230
& 0.0462 & 0.0252 & 0.0342 \\
&
& $\mathcal{L}_{\Delta\mathrm{PCC}}$
& -0.0115 & 0.0035 & 0.0004
& -0.0307 & -0.0019 & 0.0030 \\

& \multirow{2}{*}{EEGDeformer}
& $\mathcal{L}_{\mathrm{PCC}}$
& 0.0275 & 0.0554 & 0.0197
& 0.0439 & 0.0359 & 0.0271 \\
&
& $\mathcal{L}_{\Delta\mathrm{PCC}}$
& -0.0009 & 0.0045 & -0.0019
& -0.0284 & -0.0093 & 0.0003 \\
\bottomrule
\end{tabular*}
\end{table*}

\subsection{Correlation Between Accuracy and PCC}

Fig.~\ref{fig:acc_relations} illustrates the relationships between decoding accuracy and (i) attended PCC and (ii) $\Delta$PCC. The decoding accuracy demonstrates a weak to moderate correlation with the attended PCC (Fig.~\ref{fig:acc_relations}a). However, a moderate to strong correlation can be observed ($R^2>0.5$) between decoding accuracy and $\Delta$PCC (Fig.~\ref{fig:acc_relations}b). This observation is consistent with the motivation of this paper that a contrastive training objective can be proposed to improve decoding performance of DNN-based envelope decoders. 

\subsection{Impact of Different Loss Functions on ACC and PCC Difference}

To evaluate the efficacy of the proposed method, all four models are trained with the $\mathcal{L}_{\mathrm{PCC}}$ and the $\mathcal{L}_{\Delta\mathrm{PCC}}$ as objective. As shown in Table~\ref{tab:acc_pcc_merge}, models trained with the $\mathcal{L}_{\Delta\mathrm{PCC}}$ achieved higher decoding accuracy than those trained with $\mathcal{L}_{\mathrm{PCC}}$ across most datasets and window lengths. Multiple gains are statistically significant, as denoted by the asterisks in Table~\ref{tab:acc_pcc_merge}. These results indicate that optimizing DNN models with the proposed contrastive PCC loss yields improved AAD performance over the conventionally used loss.

Table~\ref{tab:acc_pcc_merge} also shows that models trained with $\mathcal{L}_{\Delta\mathrm{PCC}}$ generally yield a substantially larger PCC difference compared to those trained with $\mathcal{L}_{\mathrm{PCC}}$. Averaged across all datasets, models, and window lengths, the relative improvement is \textbf{17.84\%}, indicating that the contrastive nature of the $\mathcal{L}_{\Delta\mathrm{PCC}}$ objective effectively improves the PCC difference, which is highly associated with the decoding accuracy, as shown in Fig.~\ref{fig:acc_relations}.

\subsection{Mechanism of the proposed approach}

Although the efficacy of the proposed contrastive loss has been validated, the mechanism behind it remains further explained. A possible hypothesis is that DNN model does succeed in extracting speech envelope features from the EEG signals, but the produced results contain less discriminative information to separate the attended envelopes apart from the unattended ones.

The comparison of attended and unattended PCC in Table~\ref{tab:acc_pcc_merge} supports this explanation. With $\mathcal{L}_{\mathrm{PCC}}$, the correlation with unattended speech remains relatively high, resulting in only a limited separation between the attended and unattended conditions. In contrast, training with $\mathcal{L}_{\Delta \mathrm{PCC}}$ markedly enlarges this separation by substantially reducing the correlation with unattended speech. These results suggest that the proposed contrastive loss improves decoding performance primarily by suppressing unattended-speech correlations rather than by simply increasing attended-speech correlations.

\subsection{Discussions}

Despite the consistent improvements observed in decoding accuracy and PCC difference across several models and datasets, the benefits of the $\mathcal{L}_{\Delta\mathrm{PCC}}$ are not uniformly stable. As shown in the leave-one-trial-out experiment results, a few models obtained degraded decoding performance with our proposed $\mathcal{L}_{\Delta\mathrm{PCC}}$, such as EEGMamba on the DTU dataset, indicating that the effectiveness of the proposed $\mathcal{L}_{\Delta\mathrm{PCC}}$ is sensitive to data characteristics and model configuration.

Several factors may contribute to this variability. First, differences in window length can influence the observed improvement, as the gains become more pronounced when longer windows such as 10-second segments are used. Second, the datasets differ in language, recording protocols, and experimental conditions, which can lead to inconsistent performance across corpora. A more consistent and robust optimization goal remains the potential of better performance improvement and training stability.

\section{Conclusion}

This work provided a systematic comparison of four representative EEG regression architectures for speech-envelope reconstruction across three benchmark datasets. We introduced a contrastive loss function, $\mathcal{L}_{\Delta\mathrm{PCC}}$, designed to enhance the separation between attended and unattended speech envelopes, and evaluated it against the conventional $\mathcal{L}_{\mathrm{PCC}}$ under a unified preprocessing and leave-one-trial-out training pipeline. The $\Delta\mathrm{PCC}$ objective often achieves higher decoding accuracy and substantially larger attended–unattended PCC differences. Overall, our findings highlight the complementary strengths of correlation-based and contrastive objectives and provide practical guidance for designing regression losses in EEG-based auditory attention decoding. 

\ifcameraready
\section{Acknowledgements}
Yayun Liang and Yuanming Zhang contributed equally to this manuscript. The work of Fei Chen was supported by the National Key Research and Development Program of China (Grant No.~2023YFF1203502). The work of Jing Lu was supported by The National Natural Science Foundation (NSF12274221). The work of Yayun Liang was supported by the Nanjing University Scientific Research and Practice Innovation Program (Grant No.~KYCX25\_0161).  
\fi

\bibliographystyle{IEEEtran}
\bibliography{mybib}

\end{document}